\documentclass[preprint]{aastex}

\newcommand{\be}{\begin{equation}}
\newcommand{\ee}{\end{equation}}

\begin{document}

\title{Total to Selective Extinction Ratios and Visual Extinctions
from Ultraviolet Data}

\author{Anna Geminale\altaffilmark{1} and Piotr Popowski}
\affil{Max-Planck-Institut f\"{u}r Astrophysik,
Karl-Schwarzschild-Str.\ 1, Postfach 1317, 85741 Garching bei
M\"{u}nchen, Germany\\
E-mail: {\tt geminale, popowski@mpa-garching.mpg.de}}
\altaffiltext{1}{Ph.D. student, Department of Astronomy, 
Vicolo dell'Osservatorio, 2, I-35122 Padova, Italy} 

\begin{abstract} 
We present determinations of the total to selective extinction ratio $R_V$
and visual extinction $A_V$ values for Milky Way stars using ultraviolet
color excesses. We extend the analysis of \citet{b2} by
using non-equal weights derived from observational errors. We
present a detailed discussion of various statistical errors. In
addition, we estimate the level of systematic errors by considering
different normalization of the extinction curve adopted by
\citet{b11}. Our catalog of 782 $R_V$ and $A_V$ values and their
errors is available in the electronic form on the World Wide Web.

\end{abstract}

\section{Introduction}
\label{introduction}

The extinction curve describes how the extinction changes with the
wavelength. Extinction is due to the presence of dust grains in the
interstellar medium and its characteristics are different in a diffuse
interstellar medium as compared to a dense interstellar medium.
Thus, the knowledge of extinction curve is necessary to deredden
magnitudes and colors of astronomical objects and to understand the
physical properties of dust grains.

\citet[ hereafter CCM]{b1} 
derived a mean
extinction law (for $0.12~\mu m < \lambda < 3.5~\mu m$)
that depends on only one parameter $R_V=A_V/E(B-V)$.
They considered the sample used in the ultraviolet (UV)
extinction study of \citet{b4} based on \emph{International Ultraviolet
Explorer} extinction curves
of 45 reddened Milky Way OB stars. CCM searched for the corresponding
optical and near-infrared
(UBVRIJHKL) photometry from the literature. Finally they used the
intrinsic colors of \citet{b13} for the appropriate spectral types to
determine the extinction.
They obtained the following one-parameter family of curves that
represents the UV to infrared (IR) extinction law in terms of $R_V$:

\begin{equation}
\frac{A_{\lambda}}{A_V}=a(x)+b(x) \cdot R_V^{-1},
\label{CCMlaw}
\end{equation}
where $x=1/\lambda$, and $a(x)$ and $b(x)$ are the
wavelength-dependent coefficients.
Equation (\ref{CCMlaw}) is very powerful because it allows one to
determine the 
extinction in some spectral region based on the extinction in a different
spectral region, given that one knows $R_V$.

The $R_V$ parameter ranges from about 2.0 to about 5.5 (with a typical
value of 3.1) when one goes from diffuse
to dense interstellar medium. In this formalism $R_V$ therefore 
characterizes the dust properties in the region that
produces the extinction. Many authors used this parameter to 
study extinction, e.g.: \citet{b30} searched for a relation between
$R_V$ and other parameters that characterize the extinction curves;
\citet{b31} studied the role of $R_V$ as the main regulatory agent of
the penetration of radiation inside dark clouds; \citet{b32} discussed
different methods to deredden the data and obtained a new estimate of 
the extinction law in terms of $R_V$. \citet[ hereafter GS]{b2}
applied a $\chi^2$ minimization to compute the $R_V$ values
for a sample of stars with UV extinction data using the linear
relation (\ref{CCMlaw}). A similar method with weights was used by
 \citet{b50} to
determine $R_V$ and $A_V$ toward a sample of stars with known color
excesses in UBVRIJHKL.

Here we extend the method used by GS in order to obtain improved
$R_V$ values for the lines of sight toward a sample of stars with
known extinction data in the UV. 
The structure of this paper is the following. In \S
\ref{Th.cons.} we discuss
the theoretical basis of our $R_V$ derivation. In \S \ref{Data} we
describe our data sources. We present the results and
assess the consistency between different samples and theoretical
approaches in \S \ref{Results} . Finally in \S \ref{Conclusion} we
discuss our results and comment on the future work.

\section{Theoretical considerations}
\label{Th.cons.}

The interstellar dust grains span a wide range of sizes
from a few Angstroms to a few micrometers. In general, they reduce the
intensity of the transmitted beam by two physical processes: absorption
and scattering. The extinction is the result of these two processes.\\
The apparent magnitude $m$ of each star as a
function of wavelength may be written as
\begin{equation}
m_{\lambda,\rm red}=M_{\lambda,\rm red} + 5 \log \frac{d_{\rm
red}}{10{\rm pc}} + A_{\lambda,\rm red},
\label{mred}
\end{equation}
\begin{equation}
m_{\lambda,\rm comp}=M_{\lambda,\rm comp} + 5 \log \frac{d_{\rm
comp}}{10{\rm pc}} +
A_{\lambda,\rm comp},
\label{mcomp}
\end{equation}
where $M$, $d$ and $A$ represent absolute magnitude, distance and
total extinction, respectively, and subscripts ``red'' and
``comp'' denote 'reddened' and 'comparison' stars, respectively.
The extinction as a function of $\lambda$ may be obtained by comparing
corresponding stars paired according to spectral properties. In
principle,
the
'comparison' star should be of the same spectral
classification as the 'reddened' star, but with a negligible extinction.

If the reddened star and the comparison star have the same spectral
classification it
also means that they have very similar intrinsic spectral energy
distributions. Thus we have 
$M_{\lambda,\rm red}=M_{\lambda,\rm comp}$. We also assume that
$A_{\lambda}\equiv A_{\lambda,\rm red} \gg A_{\lambda,\rm comp}$. 
The magnitude difference obtained from equation (\ref{mred}) and
(\ref{mcomp}) is therefore:

\begin{equation}
\Delta m_{\lambda} = m_{\lambda,\rm red} - m_{\lambda,\rm comp} = 5 \log
\left(\frac {d_{\rm red}}{d_{\rm comp}} \right) + A_{\lambda}
\end{equation} 

\noindent
The quantity $5 \log (d_{\rm red}/d_{\rm comp})$ is a constant
term and may  be
eliminated by normalizing with respect to extinction
difference in two standard
wavelengths $\lambda_1$ and $\lambda_2$:

\begin{equation}
E_{\rm norm}(\lambda,\lambda_1,\lambda_2) = \frac{\Delta m_{\lambda} - 
\Delta m_{\lambda_2}}{\Delta
m_{\lambda_1}  - \Delta m_{\lambda_2}}
= \frac {A_{\lambda} - A_{\lambda_2}}{A_{\lambda_1}  - A_{\lambda_2}}
\end{equation}

Generally, the extinction curves are normalized with respect to the B
and V passbands in the \citet{b13} system:

\begin{equation}
E_{\rm norm}(\lambda,B,V)=\epsilon (\lambda-V) =
\frac{A_{\lambda}-A_{V}}{A_{B}-A_{V}} 
= \frac{E(\lambda-V)}{E(B-V)}
\end{equation} 

\noindent
where $E(\lambda-V)=A_{\lambda}-A_{V}=(m_\lambda-m_V)-(m_\lambda-m_V)_0$,
$(m_\lambda-m_V)$ is the observed color and $(m_\lambda-m_V)_0$ is the
 intrinsic color (by construction equal to the color of the 
'comparison' star).

It is possible to obtain the absolute extinction by using the total to
selective extinction ratio:

\begin{equation}
R_V=\frac{A_V}{E(B-V)}.
\end{equation}

\noindent
Then:

\begin{equation}
\epsilon(\lambda-V)=\frac{E(\lambda-V)}{E(B-V)}=
\frac{A_{\lambda}-A_{V}}{E(B-V)} 
= R_V \left \{\frac{A_{\lambda}}{A_{V}}-1 \right \}.
\label{Eps}
\end{equation}
 
\noindent
CCM, for computational reasons, divided the complete extinction curve
(equation \ref{CCMlaw}) into three wavelengths regions and fitted the
extinction law as a function of $x=(1\mu m)/\lambda$:\\
-- infrared ($0.3 \leq x \leq
1.1 $),\\
-- optical/NIR ($ 1.1  \leq x \leq
3.3$),\\
-- ultraviolet and far-ultraviolet ($ 3.3 
\leq x \leq 8.0$).\\
For every wavelength, the coefficients $a(x)$ and $b(x)$ from
equation (\ref{CCMlaw}) are fixed and given by an appropriate 
expression\footnote{In GS one of the
coefficients in their equation (2) is incorrect but the results
reported in their Table 1 suggest that they used the proper formula.}.
Observations from \emph{International Ultraviolet Explorer} cover a
range from $3.03 <x<6.45 $ ($\lambda$[$\AA$] = 1549, 1799, 2200, 2493,
and 3294), so we use the equations for the coefficients 
for the last two regions listed above.

\noindent
For $ 1.1  \leq x \leq
3.3$ and $y\equiv(x-1.82)$ we have:
\begin{eqnarray}
&a(x)&=1+0.17699y-0.50447y^2-0.02427y^3+0.72085y^4 \nonumber\\
&&~~+0.01979y^5-0.77530y^6+0.32999y^7; \nonumber\\
&b(x)&=1.41338y+2.28305y^2+1.07233y^3-5.38434y^4 \nonumber\\
&&~~-0.62251y^5+5.30260y^6-2.09002y^7.
\end{eqnarray}
For $ 3.3  \leq x \leq 8.0 $:
\begin{eqnarray}
&a(x)&=~~1.752-0.316x-0.104/[(x-4.67)^2+0.341]+F_a(x);\\
&b(x)&=-3.090+1.825x+1.206/[(x-4.62)^2+0.263]+F_b(x).
\end{eqnarray}
where:
$$
\left.
\begin{array}{rl}
F_a(x)=-0.04473(x-5.9)^2-0.009779(x-5.9)^3 \\  
F_b(x)=~~0.21300(x-5.9)^2+0.120700(x-5.9)^3\\  
\end{array}
\right\}
\mbox {if } 8 \geq x \geq 5.9
$$
\begin{equation}
F_a(x)=0=F_b(x) \qquad \mbox{otherwise}
\end{equation}
\citet{b2} compute $R_V$ values using equations
(\ref{CCMlaw}) and (\ref{Eps}) and by minimizing the quantity:
\begin{equation}
\chi^2=\sum_{i=1}^{N_{\rm bands}}\{E(\lambda_i-V) - E(B-V) \cdot
[R_V(a(x_i)-1) + b(x_i)]\}^2
\label{chi}
\end{equation}
The right side of equation (\ref{chi}) is a second order polynomial
(parabola) of $R_V$ with the minimum:
\begin{equation}
R_V=\frac{\sum_{i=1}^{N_{\rm bands}} \left\{(a(x_i)-1)\cdot
\left[ E(\lambda_i-V)/E(B-V)-b(x_i)\right]\right\}}{\sum_{i=1}^{N_{\rm bands}}(a(x_i)-1)^2}
\label{RvGS}
\end{equation}
\\
This formula is right when the errors in $E(\lambda_i-V)/E(B-V)$ are
identical for all bands. Our data (see \S \ref{Data}) have 
errors that differ from band to band, so we use an improved
$\chi^2$, weighted by the observational errors. Ducati et al.\ (2003) suggested
the following $\chi^2$ for independent minimization with respect to
$R_V$ and $A_V$:
\begin{equation}
\chi^2=\sum_{i=1}^{N_{\rm bands}} w_{\lambda_i} \left[E(\lambda_i-V)-
(a(x_i)-1)A_V- b(x_i) \frac{A_V}{R_V} \right]^2
\end{equation}
where $w_{\lambda_i}$ are the weights associated with each band. 
We use a related but different approach which stems from the fact that
in addition to UV bands we also use $E(B-V)$ as our input data. We 
normalize our color
excesses with $E(B-V)$ to form $\epsilon(\lambda-V)$. Since
$A_V = R_V \, E(B-V)$, we minimize the
following $\chi^2$ with respect to $R_V$ only: 
\begin{equation}
\chi^2=\sum_{i=1}^{N_{\rm bands}} w_{\lambda_i} \{
\epsilon(\lambda_i-V) - [R_V(a(x_i) -1)+
b(x_i)] \} ^2 ~E^2(B-V)
\label{chiweighted}
\end{equation}
Setting $w_{\lambda_i} \equiv 1/\sigma_i^2$ and minimizing equation (\ref{chiweighted}) with respect to $R_V$ we find:
\begin{equation}
R_V=\frac{\sum_{i=1}^{N_{\rm bands}} \{ (a(x_i)-1)\cdot
[\epsilon(\lambda_i-V)-b(x_i)]/ \sigma_i^2
\}}{\sum_{i=1}^{N_{\rm bands}} \{ (a(x_i)-1)^2/
\sigma_i^2 \} }
\label{Rvweighted}
\end{equation}
where: 
\begin{eqnarray}
\sigma_i^2 &\equiv&  \sigma^2 [\epsilon(\lambda_i-V)] \equiv 
\left(\frac{\partial \epsilon(\lambda_i-V)}{\partial
E(\lambda_i-V)} \sigma[E(\lambda_i-V)]\right)^2 + \left
(\frac{\partial \epsilon(\lambda_i-V)}{\partial
E(B-V)} \sigma[E(B-V)]\right)^2  \nonumber\\
&=&
\left(\frac{E(\lambda_i-V)}{E(B-V)} \right)^2 \left[ \left(
\frac{\sigma[E(\lambda_i-V)]}{E(\lambda_i-V)}\right)^2 +
\left(\frac{\sigma[E(B-V)]}{E(B-V)} \right)^2 \right]
\label{sigma}
\end{eqnarray}
and 
\begin{eqnarray}
\sigma^2[E(\lambda_i-V)] &\equiv& \sigma^2[(m_{\lambda_i}-m_V) -
(m_{\lambda_i}-m_V)_0] \nonumber\\
&=&\sigma^2[m_{\lambda_i}]+\sigma^2[m_V]+\sigma^2_{i,{\rm mismatch}}
\label{sigmaE}
\end{eqnarray}
The error terms on the right side of equation (\ref{sigmaE}) are
described in Table~\ref{table1}.
In equation (\ref{sigma}) we assumed for simplicity that the errors in
$E(\lambda_i-V)$ and $E(B-V)$ are independent. However, the values of
$\epsilon(\lambda-V)$ and their errors for different bands are
not independent. To get a good idea about the errors in $R_V$ we
compute them in two ways. First, we calculate the maximum error in
$R_V$, which is the straight sum of errors coming from different sources:
\begin{equation}
\sigma_{\rm max} (R_V) \equiv \sum_{j=1}^{N_{\rm bands}} \left
|\frac{\partial R_V}{\partial
\epsilon(\lambda_j-V)} \right| \cdot \sigma_j
= \frac{1}{\sum_{i=1}^{N_{\rm bands}}
(a(x_i)-1)^2/\sigma_i^2} \cdot \sum_{j=1}^{N_{\rm bands}} \left | 
\frac{a(x_j)-1}{\sigma_j} \right|
\label{errabs}
\end{equation}
Then we obtain the error in quadrature which would properly describe
total uncertainty if the errors from different sources were uncorrelated:

\begin{eqnarray}
\sigma_{\rm quad} (R_V) &\equiv& 
\sqrt{\sum_{j=1}^{N_{\rm bands}} \left [ \left
(\frac{\partial R_V}{\partial
\epsilon(\lambda_j-V)} \right)^2 \cdot \sigma_j^2 \right]} \nonumber\\
&=&\frac{1}{\sum_{i=1}^{N_{\rm bands}}
[(a(x_i)-1)^2/\sigma_i^2]} \cdot \sqrt{ \sum_{j=1}^{N_{\rm bands}} \left (
\frac{a(x_j)-1}{\sigma_j} \right)^2}
\label{quaerr}
\end{eqnarray}
Neither description (\ref{errabs}) nor (\ref{quaerr}) is strictly
correct: the real error in $R_V$ lies likely between these two
estimates.  

By definition:

\begin{equation}
A_V \equiv R_V \, E(B-V) = 
\frac{\sum_{i=1}^{N_{\rm
bands}}(a(x_i)-1)(E(\lambda_i-V)-b(x_i)E(B-V))/\sigma_i^2}{\sum_{i=1}^{N_{\rm
bands}}(a(x_i)-1)^2/\sigma_i^2},
\label{Avformula}
\end{equation}
where the second equality is a consequence of equation (\ref{Rvweighted}).
Therefore the maximum error in $A_V$ is given by\footnote{We note that
the analysis of GS is equivalent to ours if
weights meet the following conditions: 
$\bigwedge_ {i \in [\overline{1,N_{\rm bands}}]} \frac{1}{\sigma_i^2} = C$,
where $C$ is a constant. From equation~(\ref{sigma}) we conclude
that this condition is in conflict with 
$\sigma [E(\lambda_i-V)]=\sigma [E(B-V)]$ assumed by GS. 
If we ignore this conflict and force both conditions, then:
$$
\sigma_{\rm max} (A_V)
= \left [
\frac{\sum_{j=1}^{N_{\rm bands}} |a(x_j)-1|}{\sum_{i=1}^{N_{\rm
bands}}
(a(x_i)-1)^2} +
\frac{| \sum_{i=1}^{N_{\rm
bands}}(a(x_i)-1)\cdot(-b(x_i))|}{\sum_{i=1}^
{N_{\rm bands}}(a(x_i)-1)^2} \right ] \cdot
\sigma [E(B-V)] 
$$
The results reported in Table 1 of GS suggest that they used the
above formula rather than their equation (8).}:
\begin{eqnarray}
\sigma_{\rm max} (A_V) &\equiv&
\sum_{j=1}^{N_{\rm bands}} \left| 
\frac{\partial A_V}{\partial
E(\lambda_j-V)} \right | \sigma [E(\lambda_j-V)] + \left |
\frac{\partial A_V}{ \partial E(B-V)} \right| \sigma [E(B-V)] \nonumber\\
&=&
\frac{1}{\sum_{i=1}^{N_{\rm bands}} (a(x_i)-1)^2/\sigma_i^2}
\left[ \sum_{j=1}^{N_{\rm bands}} \left | \frac{(a(x_j)-1)}{\sigma_j^2}\right|
\sigma[E(\lambda_j-V)] \right. \nonumber\\
&+& \left. \left| \sum_{i=1}^{N_{\rm
bands}}\frac{(a(x_i)-1)(-b(x_i))}{\sigma_i^2} \right| \sigma[E(B-V)] \right]
\label{Averrmax}
\end{eqnarray}
The error in quadrature is given by:
\begin{eqnarray}
\sigma_{\rm quad} (A_V) &\equiv& \sqrt{\sum_{j=1}^{N_{\rm bands}}
\left ( \frac{\partial A_V}{\partial E(\lambda_j-V)} \right )^2 
\sigma^2[E(\lambda_j-V)]
+ \left ( \frac{\partial A_V}{\partial E(B-V)} \right )^2 \sigma^2[E(B-V)]}
\nonumber\\
&=&
\frac{1}{\sum_{i=1}^{N_{\rm bands}}
(a(x_i)-1)^2/\sigma_i^2}
\left[\left(\sum_{j=1}^{N_{\rm bands}} \frac{(a(x_j)-1)}{\sigma_j^2}
\right)^2\sigma^2[E(\lambda_j-V)]\right. \nonumber\\
&+&
\left.\left(\sum_{i=1}^{N_{\rm
bands}}\frac{(a(x_i)-1)(-b(x_i))}{\sigma_i^2 }
\right)^2\sigma^2[E(B-V)] \right]^{1/2}
\label{sigmaquadAv}
\end{eqnarray}

\section{Data}
\label{Data}

\noindent
We use the data taken from the \citet{b5} catalog of ultraviolet color
excesses 
\begin{equation}
E(\lambda-V)=(m_{\lambda}-m_V)-(m_{\lambda}-m_V)_0
\label{extinction}
\end{equation}
for stars of spectral types B7 and earlier.
The UV measurements are taken from \emph{Astronomical Netherlands Satellite}
(ANS) data \citep{b6} and consist of observations in five
UV channels with central wavelengths:
$\lambda=$ 1549, 1799, 2200, 2493, and 3294$\AA$. 

\placetable{table1}

The sources of the data used to obtain $E(\lambda - V)$ as given by 
equation (\ref{extinction})
and their errors are listed in Table \ref{table1}. We also
consider another type of error: a `mismatch error', which is caused by
the fact that the reddened star and the comparison star may have slightly
different colors. 
\citet{b8} give in their section 2c, Table 1B ultraviolet color excess
errors which include errors associated with spectral type
misclassification (mismatch error) and errors in the intrinsic
colors. We adopt their values for this total additional source
of error, and we record them under the name of mismatch error.

\begin{figure}
\begin{center}
\includegraphics[width=150mm]{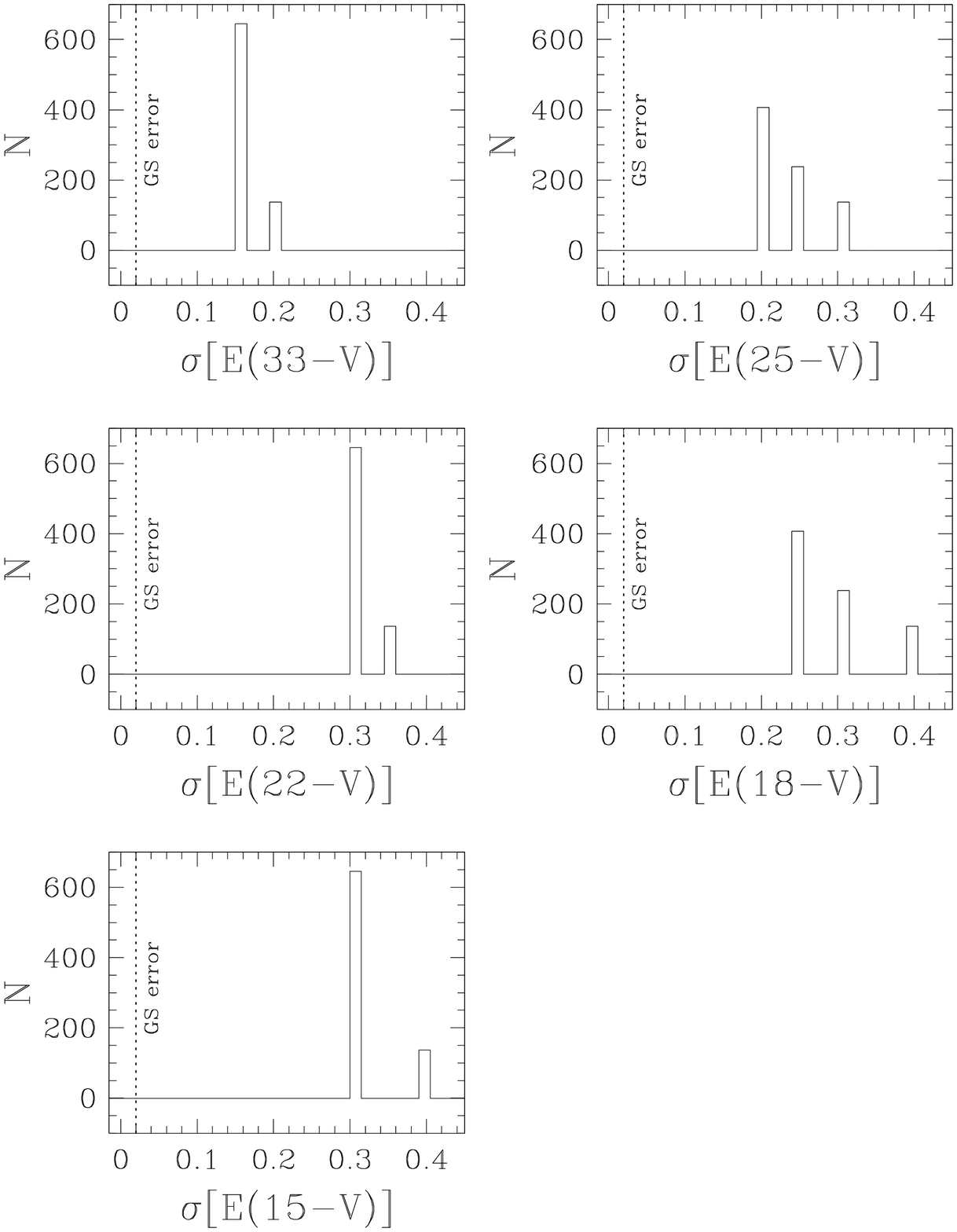}
\caption{Histograms of the errors in the $E(\lambda_i-V)$ for all 782
stars. The vertical lines represent the values
adopted by GS for the errors in $E(\lambda_i-V)$.}
\label{figure1}
\end{center}
\end{figure}

\placefigure{figure1}

Figure \ref{figure1} shows the histograms of the $E(\lambda_i-V)$ errors 
for the five ultraviolet bands which we obtain using 
equation (\ref{sigmaE}). The errors are completely dominated by the
mismatch errors which results in a few spikes observed in each panel.
The errors adopted by GS and marked by vertical lines are shown for
comparison.

From the \citet{b5} catalog we exclude some
lines of sight using the same method of selection as GS. It means that
we exclude the lines of sight that have $E(B-V)<0.1$, and the ones
with $E(\lambda-V)/E(B-V)>8$. This selection results in 923 lines of
sight considered previously by GS. In addition, we also exclude those
stars that do not have
spectral type classification, because for them we are not able to
assign the mismatch errors. This last cut reduces the number of
lines of sight we consider to 782.

\begin{figure}
\begin{center}
\includegraphics[width=150mm]{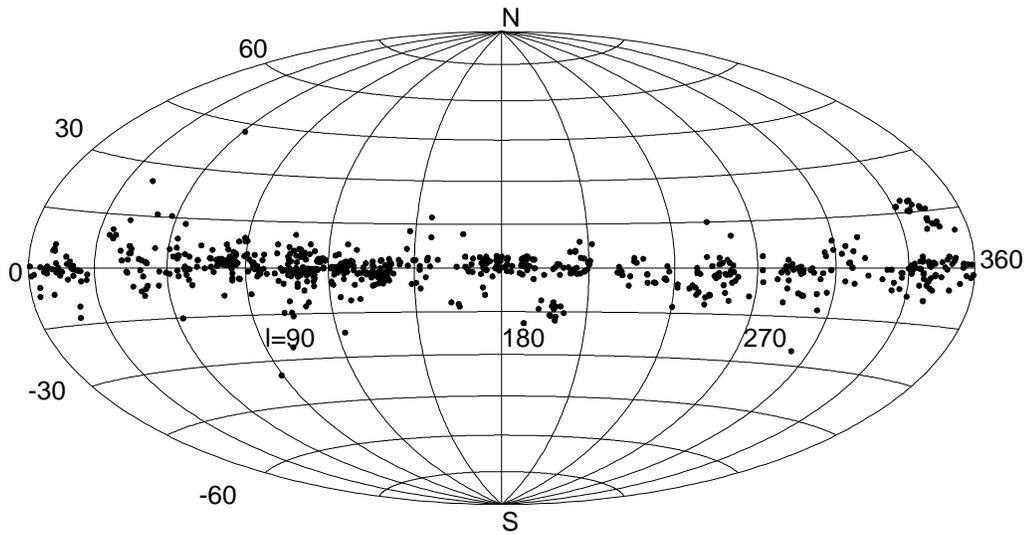}
\caption{Sky distribution of the stars in our sample (galactic coordinates).}
\label{figure2}
\end{center}
\end{figure}

\placefigure{figure2}

Figure \ref{figure2} shows the sky positions of the stars in our
sample. The sample contains stars of spectral type B7 and earlier and this is
the reason for which almost all the stars lie in the Galactic plane at
low latitudes.\\

\section{Results}
\label{Results}

\subsection{The catalog of $R_V$ and $A_V$ values}

\placetable{table2}

By using the method described in \S \ref{Th.cons.} we compute
for our sample the $R_V$ and $A_V$ values listed in Table \ref{table2}. Here
we present only the first 20 objects from our sample. The complete table is
available in electronic form and on the World Wide Web\footnote{See
{\tt http://dipastro.pd.astro.it/geminale}}. In the first
column we list the names of stars, in the second and third the
galactic coordinates, then the 
$E(B-V)$ values taken from \citet{b5} catalog; in the remaining columns
we list the $R_V$ and $A_V$ values with their errors obtained using 
formulae (\ref{Rvweighted})--(\ref{sigmaquadAv}).\\

\begin{figure}
\begin{center}
\includegraphics[width=150mm]{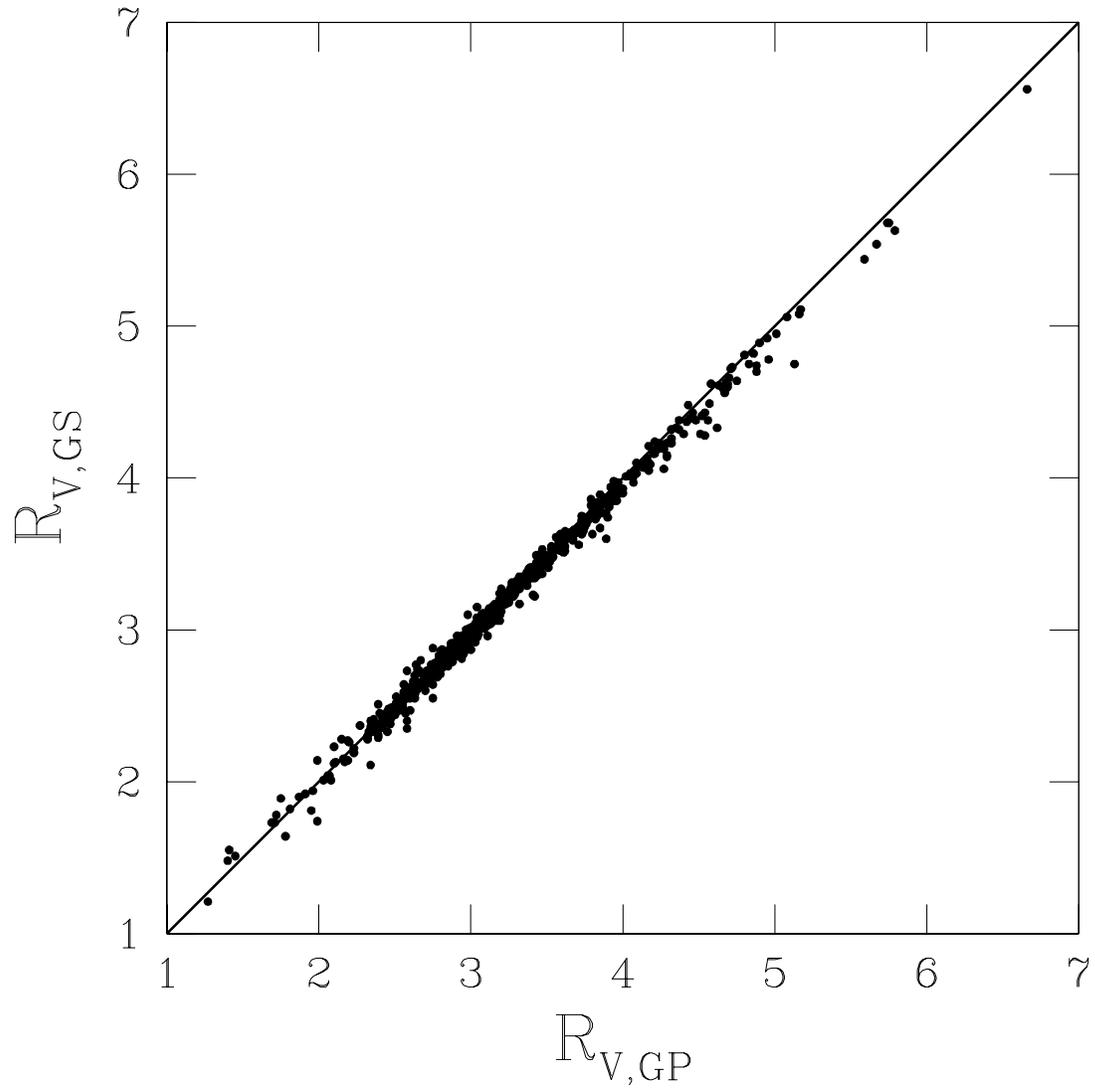}
\caption{Comparison between the $R_V$ values obtained by GS and the $R_V$
values obtained using our method. The line shows the 1-to-1 relation.}
\label{figure3}
\end{center}
\end{figure}

\placefigure{figure3}

Our determination of $R_V$ values with their errors is made using the GS method
improved through the consistent treatment of observational errors. Figure
\ref{figure3} shows the good agreement between the $R_V$
values obtained with GS unweighted method and the weighted method
applied here. 
In our case, the $R_V$ values are not so different
between the
two methods because our adopted errors in $\epsilon(\lambda_i-V)$ are 
of the same order in the five UV wavelengths. However, it's important to
notice that typical
errors in $R_V$ are very different between the two methods 
(mostly due to the mismatch errors considered here). \\

\begin{figure}
\begin{center}
\includegraphics[width=150mm]{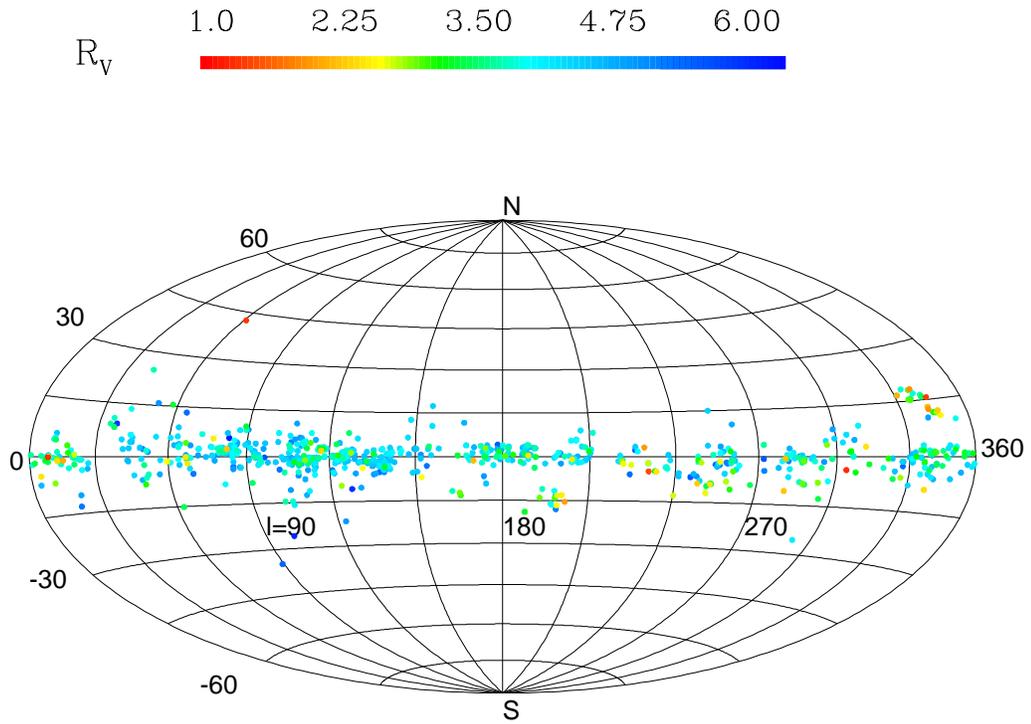}
\caption{Sky distribution of the $R_V$ values (galactic coordinates).}
\label{figure4}
\end{center}
\end{figure}

\placefigure{figure4}

\begin{figure}
\begin{center}
\includegraphics[width=80mm]{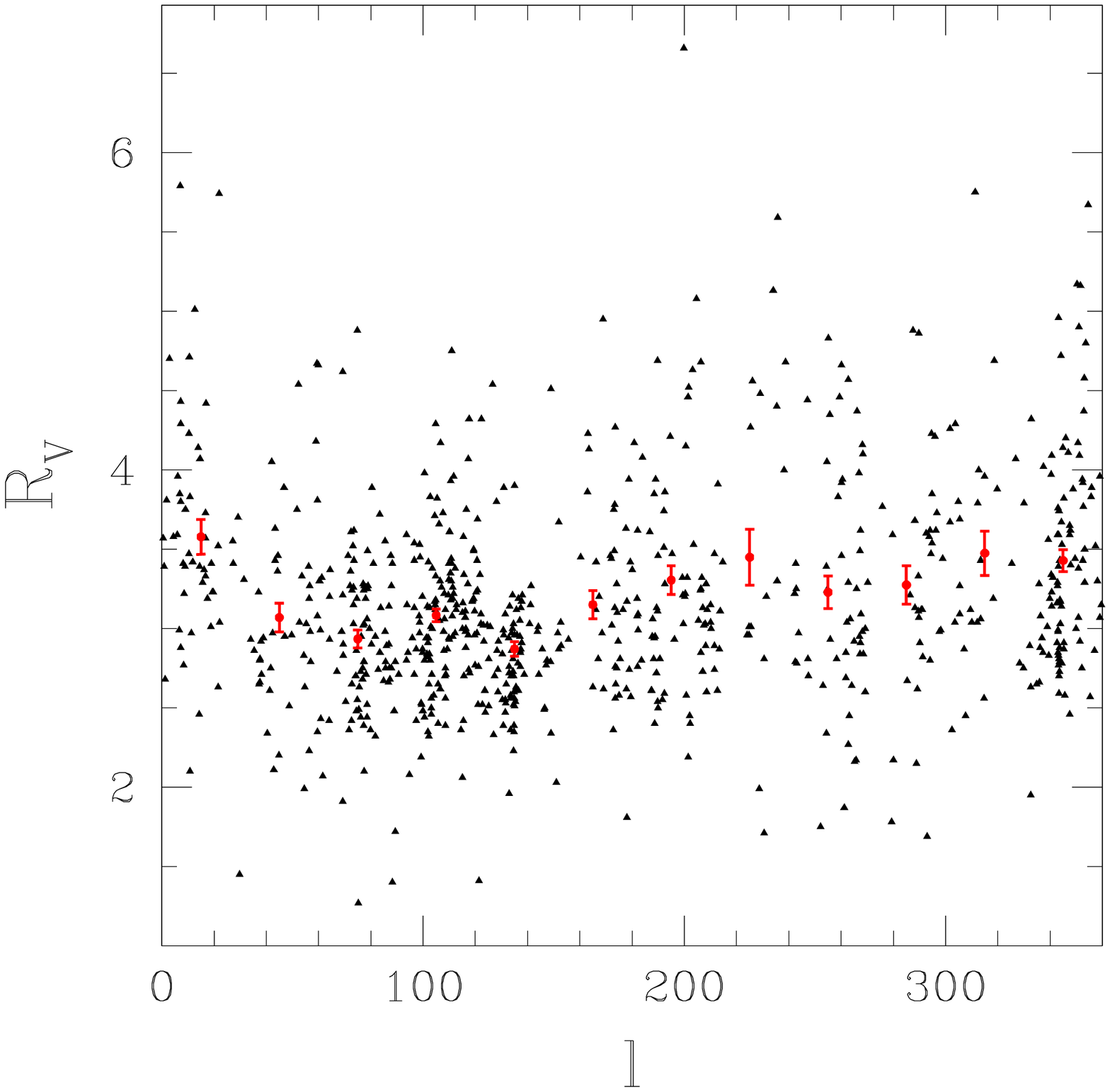}
\includegraphics[width=80mm]{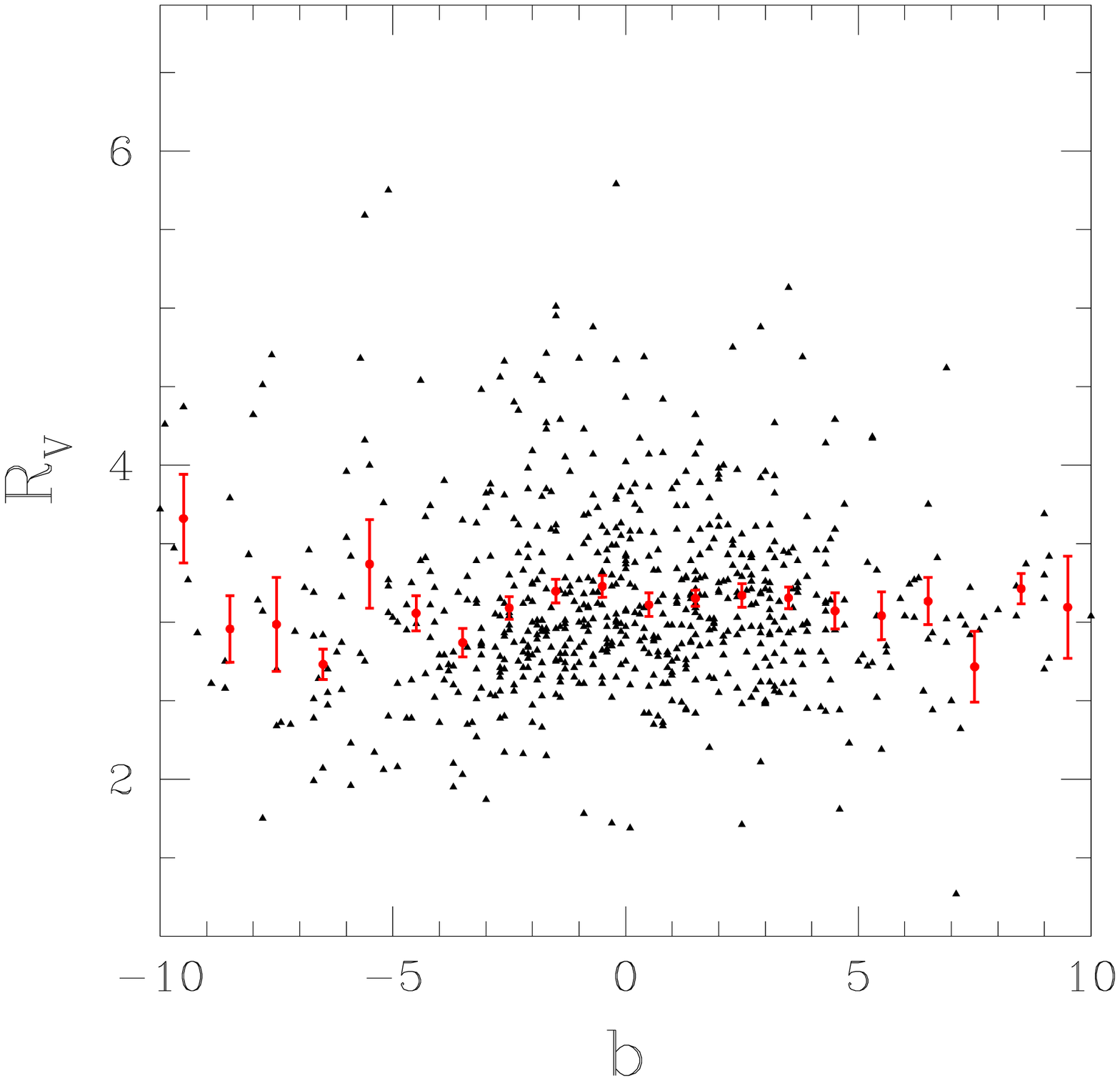}
\caption{$R_V$ values as a function of galactic coordinates. The
circular red points represent the unweighted mean values of $R_V$ in different
coordinate bins with the rms error bars represented by the vertical
lines. The galactic longitudes and latitudes are binned every $30^\circ$
and $1^\circ$, respectively.}
\label{figure5}
\end{center}
\end{figure}

\begin{figure}
\begin{center}
\includegraphics[width=150mm]{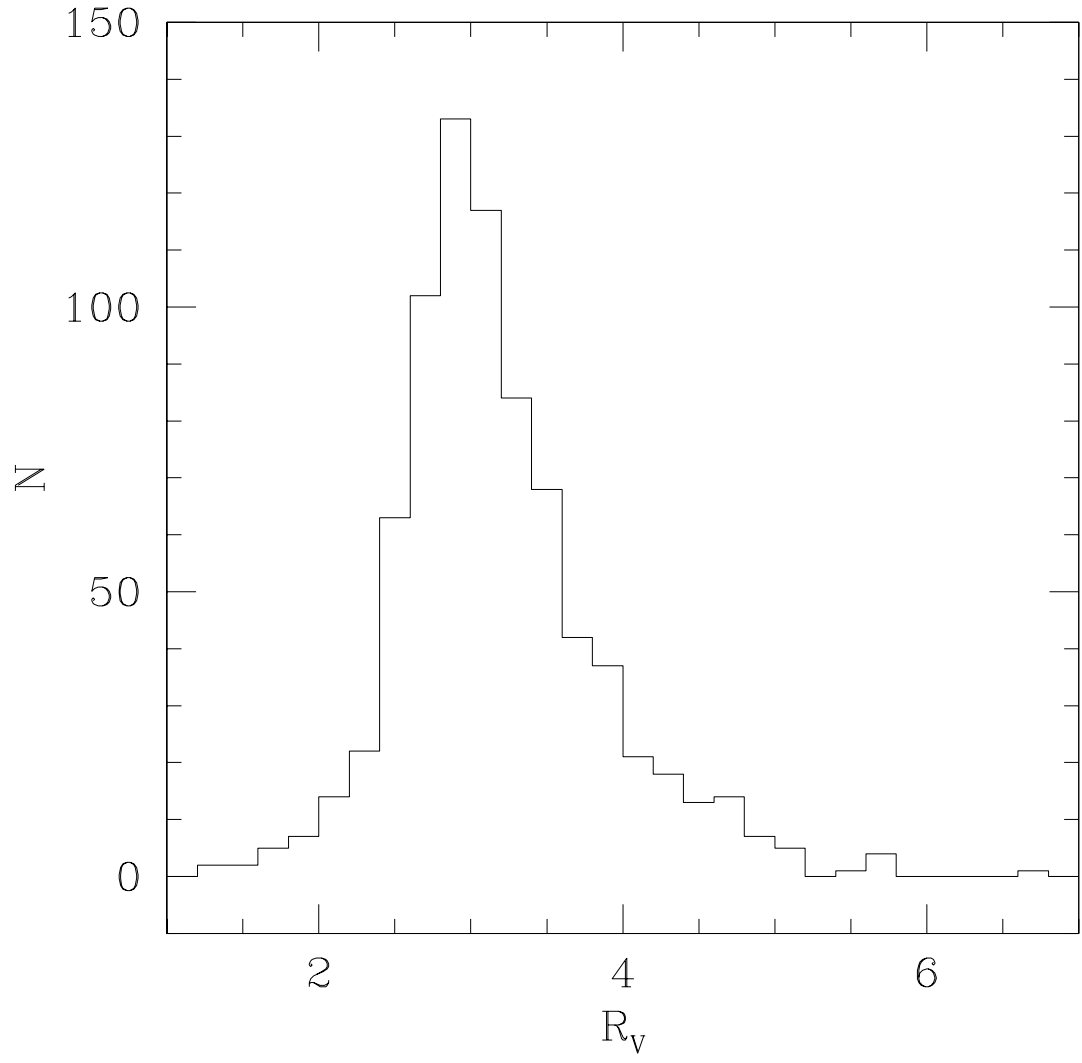}
\caption{Histogram of the $R_V$ values computed in this paper and
listed in Table \ref{table2}.}
\label{figure6}
\end{center}
\end{figure}

Figure \ref{figure4} shows the same points as in Figure
\ref{figure2}, but now different colors mark different values of
$R_V$ with $R_V$ increasing from red to blue. As
expected most lines of sight have $R_V$ of about 3.1.
This may be also seen in Figure \ref{figure5} which shows $R_V$
values as a function of galactic coordinates. The circular red points
are the mean values of $R_V$ for the data binned every $30^{\circ}$ and
every $1^{\circ}$ for the Galactic longitude and latitude, respectively. These
mean values do not differ a lot one from another  but some sky anisotropy
is also quite apparent. Figure \ref{figure6} shows the histogram of the
$R_V$ values derived here. The weighted mean of $R_V$ values is 
$3.13 \pm 0.02$.

\placefigure{figure5}
\placefigure{figure6}

\subsection{Analysis of systematic errors}

We consider systematic errors in extinction curve
determination that can result from using biased $E(B-V)$ values. 
To this aim we use the Wegner's (2002) calibration of $E(B-V)$
to estimate the effect of the calibration change on the value
of $R_V$.
Usually, the extinction curve is expressed in terms of a color excesses to
$E(B-V)$ ratio. Since $R_V$ value depends on this ratio (see equation
\ref{Eps} or \ref{Rvweighted}), the adopted $E(B-V)$ calibration will influence
it.
Since we do not know which set of $E(B-V)$ values is more appropriate
[\citet{b5} or \citet{b11}], the
difference in obtained $R_V$ values will be a good indicator of
a possible systematic error in $R_V$.

\citet{b11} made a catalog of interstellar extinction curves of OB
stars. He used the UV data from \citet{b6}, but differently from
\citet{b5} who used the data sources described in Table~\ref{table1}, 
he took the visual magnitudes and spectral
classification of O and B stars from the SIMBAD database.
The maximum error in $E(B-V)$ adopted by \citet{b11} is 0.04 mag; the error in
$m_{\lambda}$ and $m_V$ is 0.01, and he obtains the intrinsic colors 
using the `artificial' standard method by
\citet{b12}, who found a linear relation between $(m_{\lambda}-m_V)$
and $(m_B-m_V)$ and used the coefficients of this relation also to compute
the linear relation between the intrinsic colors. This method 
improves the accuracy of the intrinsic colors based on ANS photometry.

There are 190 stars that \citet{b11} has in common with
our sample. For these stars we compute the $R_V$ values using 
formula  (\ref{Rvweighted}) weighted by the observational errors given by
\citet{b11}. 

\begin{figure}
\begin{center}
\includegraphics[width=80mm]{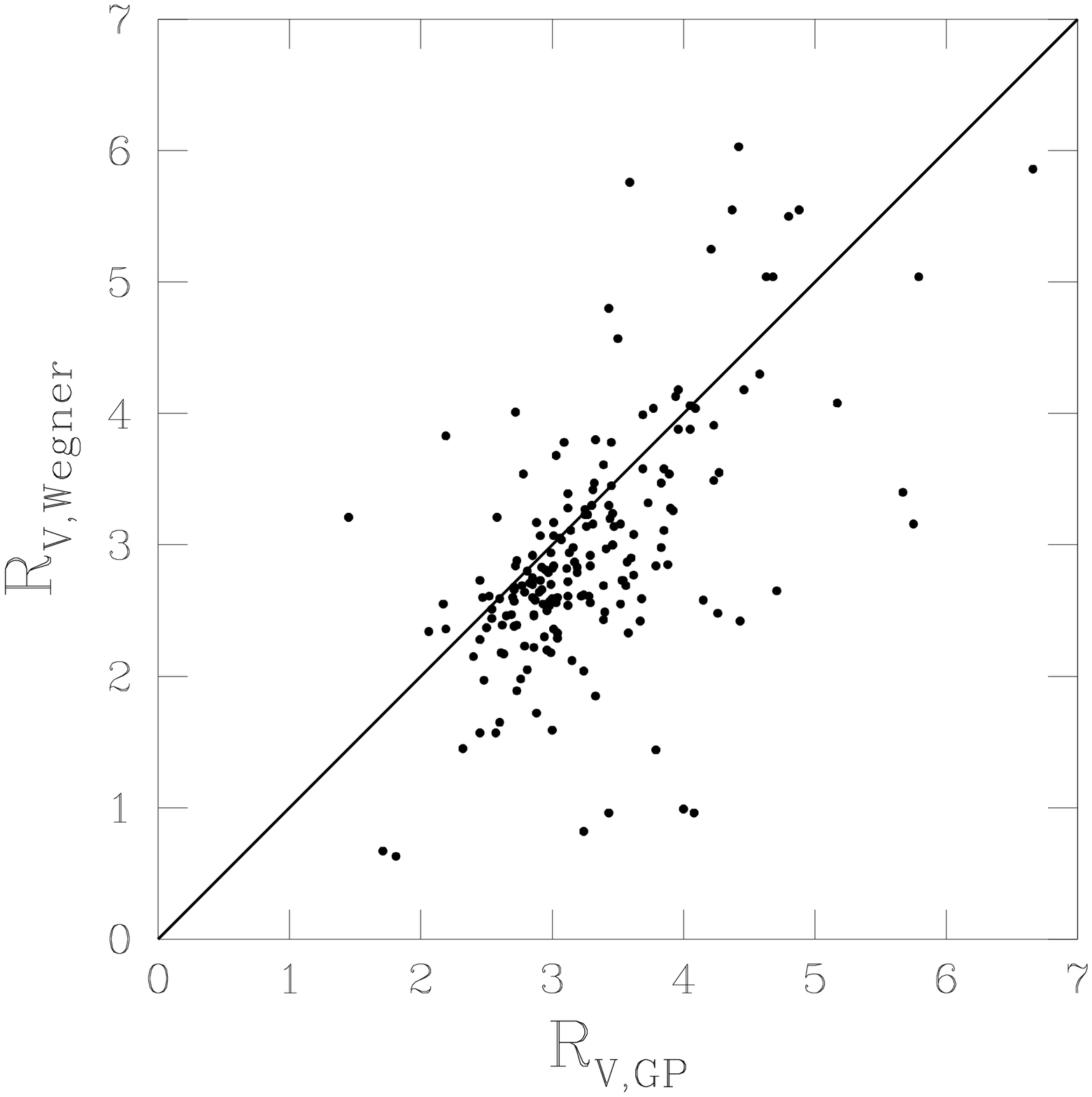}
\includegraphics[width=80mm]{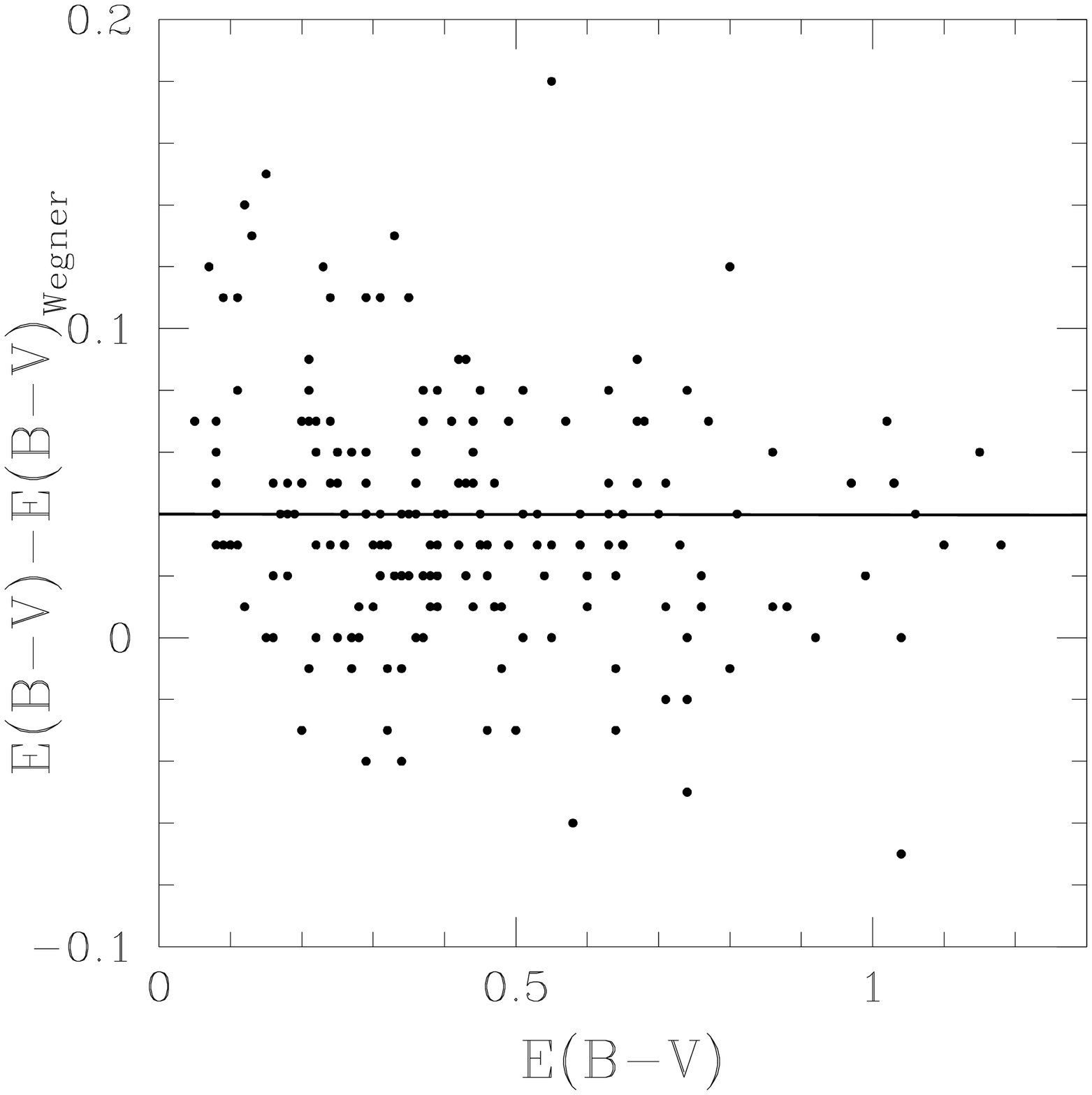}
\caption{The left panel shows $R_V$ values obtained from
equation (\ref{Rvweighted}) using the Wegner's (2002) UV data versus the
$R_V$ values computed with the same formula, but using our primary
data. The line shows a 1-to-1 relationship. The
right panel shows the comparison between two different
calibrations of $E(B-V)$. The line marks the average difference level.}
\label{figure7}
\end{center}
\end{figure}

\placefigure{figure7}

Figure \ref{figure7} shows the difference in the $R_V$ values given by
different calibrations\footnote{The linear
relation between the two set of $R_V$ values is: $R_{V,\rm Wegner}=(-1.856
\pm 0.483)+(1.530 \pm 0.148) \cdot R_{V,{\rm GP}}$.}. 
The main effect on $R_V$ comes from the fact that the $E(B-V)$ values
from two calibrations differs on average by 0$^m$.04 in the sense that
\citet{b11} color
excesses are smaller than the ones used in our primary determination.\\
Table \ref{table3} reports results for the lines of sight in common between
Wegner's (2002) sample and our sample. The complete table is given in
the electronic form on the World Wide Web\footnote{See {\tt http://dipastro.pd.astro.it/geminale}}.  The first column lists
stellar designations, the second and third report the galactic
coordinates; in the fourth column we list the $E(B-V)$ values taken
by \citet{b5}; the fifth column gives our $R_V$ values and the sixth 
their maximum errors; the
seventh contains the $E(B-V)$ values used by \citet{b11}, the
eighth provides the $R_V$ values and the ninth their errors computed
with our method and using the Wegner's (2002) ultraviolet data.

\placetable{table3}

\section{Conclusion}
\label{Conclusion}
Using ultraviolet color excesses we find $R_V$ and
$A_V$ values
and their errors for a sample of 782 lines of sight. We extend the
analysis of \citet{b2} by considering various sources of statistical
and systematic errors. In a treatment related to the one by  Ducati et
al. (2003), we introduce the weights associated with the errors in
each UV band to our $\chi^2$ minimization procedure. 
We explicitly give all the 
formulae we use to compute the $R_V$ and $A_V$ values and their
errors. We compute the maximum errors and the errors in quadrature for
$R_V$ and $A_V$ taking into account mismatch errors that affect the 
color excesses to the largest extent.
We present the sky distribution of $R_V$ values and show their
behavior as a function of galactic coordinates.
Finally, we emphasize how $R_V$ values change with different
calibrations of $E(B-V)$.
Since $R_V$ value may characterize entire extinction curves, extending our
study into wavelength regions beyond ultraviolet will provide a 
check on the universality of CCM law in various parts of the
spectrum. We discuss this issue in the forthcoming paper.

\acknowledgments

We are grateful to Gregory Rudnick for his very careful
reading of the original version of this manuscript and a number
of helpful suggestions. We thank Paola Mazzei and Guido Barbaro
for their comments. 
AG acknowledges the financial support from EARASTARGAL fellowship
at Max-Planck-Institute for Astrophysics, where this work has
been completed.

\clearpage

\clearpage

\begin{deluxetable}{clcp{6.5cm}}
\tablecolumns{4}
\tablewidth{0pc}
\tablecaption{Data sources and adopted errors.}
\tablehead{
\colhead{Quantity} &\colhead{Reference} &\colhead{Error range} 
&\colhead{Comments}}
\startdata
$m_V$ &\citet{b9} &0$^m$.04 &\small{We adopt the same conservative
errors estimate for all stars.}\\
\hline
$(m_B-m_V)$ &\citet{b9} &0$^m$.015 &{We adopt the same conservative
errors estimate for all stars.}\\
\hline
$(m_B-m_V)_0$ &\citet{b10} &0$^m$.02 &\\
\hline
$m_{\lambda}$ &\citet{b6} &0$^m$.001-0$^m$.218 &\small{The error is
given for every wavelength band and every line of sight. Typical
errors are of the order of tens of millimags.}\\
\hline
$(m_{\lambda}-m_V)_0$ &\citet{b70} & $--$ &\small{The error is included in the
mismatch error.}\\
\hline
$E(\lambda-V)$ &\citet{b5} &0$^m$.15-0$^m$.40 &\small{The error given here
represents only the
mismatch error $\sigma_{\rm mismatch}$ and it depends on wavelength 
band and on the
spectral type classification of the stars \citep{b8}. 
Although the mismatch error is the dominant contributor, the total
error in $E(\lambda-V)$ is expressed by equation (\ref{sigmaE}).}
\enddata
\label{table1}
\end{deluxetable}

\clearpage 

\begin{deluxetable}{lccccccccc}
\tablecolumns{10}
\tabletypesize{\small}
\tablewidth{0pc}
\tablecaption{$R_V$ and $A_V$ values with their errors.}
\tablecomments{
Columns:
[1]  star identification number,
[2]  galactic longitude $l$,
[3]  galactic latitude $b$,
[4]  color excess $E(B-V)$,
[5]  total to selective extinction ratio $R_V$,
[6]  error in $R_V$ obtained with equation (\ref{errabs}),
[7]  error in $R_V$ obtained with equation (\ref{quaerr}),
[8]  visual extinction $A_V$,
[9]  error in $A_V$ obtained with equation (\ref{Averrmax}),
[10] error in $A_V$ obtained with equation (\ref{sigmaquadAv}).
}
\tablehead{
\colhead{name} &{$l$} &{$b$} &{$E(B-V)$} &{$R_V$} &{$\sigma_{\rm max}(R_V)$} 
&{$\sigma_{\rm quad}(R_V)$} &{$A_V$}
&{$\sigma_{\rm max}(A_V)$} &{$\sigma_{\rm quad}(A_V)$}}
\startdata
BD-84617    &37.0  &8.4  &1.22   &3.23   &0.24   &0.12   &3.94   &0.47
&0.34\\
BD-84634    &38.0  &7.4  &1.22   &2.92   &0.25   &0.12   &3.56   &0.48
&0.34\\
BD-11471    &213.4 &1.4  &0.74   &2.87   &0.44   &0.21   &2.13   &0.50
&0.36\\
BD+233762   &60.3  &$-$0.3\phm{$-$} &1.05   &3.30   &0.30   &0.15
&3.47   &0.50 &0.36\\
BD+23771    &37.2  &$-$1.4\phm{$-$} &0.93   &2.65   &0.35   &0.17
&2.46   &0.50 &0.36\\
BD+243893   &61.3  &$-$0.5\phm{$-$} &0.65   &3.32   &0.45   &0.22
&2.16   &0.47 &0.34\\
BD+341054   &173.4 &$-$0.2\phm{$-$} &0.49   &3.78   &0.59   &0.28
&1.85   &0.47 &0.34\\
BD+341059   &173.0 &0.2  &0.49   &3.75   &0.59   &0.29   &1.84   &0.47
&0.34\\
BD+341150   &175.1 &2.4  &0.44   &2.58   &0.69   &0.33   &1.13   &0.47
&0.34\\
BD+341162   &175.5 &2.6  &0.36   &2.81   &0.84   &0.40   &1.01   &0.47
&0.34\\
BD+343631   &69.2  &6.9  &0.13   &4.62   &2.20   &1.05   &0.60   &0.48
&0.34\\
BD+354258   &77.2  &$-$4.7\phm{$-$} &0.29   &2.39   &1.07   &0.51
&0.69   &0.48  &0.34\\
BD+361261   &174.1 &4.3  &0.52   &2.75   &0.62   &0.30   &1.43   &0.50
&0.36\\
BD+363882   &73.5  &2.2  &0.64   &3.43   &0.50   &0.24   &2.19   &0.50
&0.36\\
BD+364145   &77.5  &$-$2.0\phm{$-$} &0.96   &2.79   &0.31   &0.15   &2.68   &0.47
&0.34\\
BD+373945   &77.3  &$-$0.2\phm{$-$} &1.07   &3.24   &0.30   &0.14   &3.46   &0.50
&0.36\\
BD+374092   &80.2  &$-$4.2\phm{$-$} &0.55   &2.90   &0.59   &0.28   &1.60   &0.50
&0.36\\
BD+391328   &169.1 &3.6  &0.88   &2.62   &0.37   &0.18   &2.30   &0.50
&0.36\\
BD+404179   &79.0  &1.2  &0.88   &3.27   &0.34   &0.16   &2.87   &0.47
&0.34\\
BD+421286   &166.1 &4.3  &0.56   &3.12   &0.53   &0.26   &1.75   &0.47
&0.34
\enddata
\label{table2}
\end{deluxetable}

\clearpage

\begin{deluxetable}{l c c c c c c c c c c}
\tablecolumns{11}
\tabletypesize{\small}
\tablewidth{0pc}
\rotate
\tablecaption{Comparison between UV-based $R_V$ values obtained using different
calibrations of $E(B-V)$.}
\tablecomments{
Columns:
[1]  star identification number,
[2]  galactic longitude $l$,
[3]  galactic latitude $b$,
[4]  color excess $E(B-V)$ taken from Savage et al.\ (1985),
[5]  total to selective extinction ratio $R_V$ computed with the calibration from Savage et al. (1985),
[6]  error in $R_V$ from column 5 obtained with equation (\ref{errabs}),
[7]  error in $R_V$ from column 5 obtained with equation (\ref{quaerr}),
[8]  color excess $E(B-V)$ taken from Wegner (2002),
[9]  total to selective extinction ratio $R_V$ computed with the calibration from Wegner (2002),
[10] error in $R_V$ from column 9 obtained with equation (\ref{errabs}),
[11] error in $R_V$ from column 9 obtained with equation (\ref{quaerr}).
}
\tablehead{
\colhead{name} &{$l$} &{$b$} &{$E(B-V)_{{\rm GP}}$} &{$R_{V,{\rm GP}}$} 
&{$\sigma_{\rm max}[R_{V,{\rm GP}}]$} &{$\sigma_{\rm quad}[R_{V,{\rm GP}}]$} &{$E(B-V)_{\rm Wegner}$} &{$R_V$}
&{$\sigma_{\rm max}[R_{V,
{\rm Wegner}}]$} &{$\sigma_{\rm quad}[R_{V,
{\rm Wegner}}]$}}
\startdata
HD1544    &119.3    &$-$0.6\phm{$-$}  &0.44   &3.19   &0.73   &0.35 &0.37     &2.79
&0.67 &0.31\\   
HD2083    &120.9    & 9.0  &0.29   &3.69   &1.01   &0.48 &0.26     &3.99
&0.70 &0.33\\   
HD2905    &120.8    & 0.1  &0.33   &3.24   &1.20   &0.57 &0.30     &0.82
&1.05 &0.49\\     
HD7252    &125.7    &$-$1.9\phm{$-$}  &0.35   &3.01   &0.86   &0.41 &0.32     &3.17
&0.68 &0.32\\    
HD12867   &133.0    &$-$3.7\phm{$-$}  &0.41   &2.72   &0.74   &0.35 &0.38     &2.84
&0.63 &0.30\\     
HD13969   &134.5    &$-$3.8\phm{$-$}  &0.56   &2.73   &0.54  &0.26  &0.54     &2.88
&0.46 &0.21\\     
HD14092   &134.7    &$-$4.1\phm{$-$}  &0.49   &2.52   &0.62  &0.30  &0.46     &2.61
&0.55 &0.26\\     
HD14250   &134.8    &$-$3.7\phm{$-$}  &0.58   &2.60   &0.52  &0.25  &0.55     &2.59
&0.49 &0.23\\    
HD14357   &135.0    &$-$3.9\phm{$-$}  &0.56   &3.90   &0.56  &0.27  &0.49     &3.28
&0.54 &0.25\\     
HD14818   &135.6    &$-$3.9\phm{$-$}  &0.48   &2.63   &0.84  &0.40  &0.46     &2.17
&0.68 &0.31\\     
HD14947   &135.0    &$-$1.8\phm{$-$}  &0.77   &2.96   &0.39  &0.19  &0.76     &2.50
&0.37 &0.17\\     
HD14956   &135.4    &$-$2.9\phm{$-$}  &0.89   &2.54   &0.45  &0.22  &0.88     &2.44
&0.36 &0.17\\     
HD16429   &135.7    & 1.1  &0.92   &3.12   &0.35  &0.17  &0.86     &2.72
&0.33 &0.16\\     
HD17114   &137.3    &$-$0.3\phm{$-$}  &0.76   &2.85   &0.40  &0.19  &0.73     &2.92
&0.35 &0.16\\     
HD17603   &138.8    &$-$2.1\phm{$-$}  &0.92   &2.71   &0.44  &0.21  &0.92     &2.66
&0.31 &0.15\\    
HD18352   &137.7    & 2.1  &0.48   &2.91   &0.63 &0.30  &0.45     &3.07
&0.52 &0.24\\     
HD24431   &148.8    &$-$0.7\phm{$-$}  &0.69   &2.79   &0.44  &0.21 &0.65     &2.64
&0.41 &0.19\\     
HD24912   &160.4    &$-$13.1\phm{$-1$} &0.29   &3.45   &1.36 &0.65  &0.26     &3.78
&0.75 &0.36\\     
HD30614   &144.1    &14.0\phm{1}  &0.30   &3.01   &1.33 &0.63  &0.34     &3.07
&0.74 &0.35\\     
HD34078   &172.1    &$-$2.3\phm{$-$}  &0.52   &3.44   &0.57 &0.27  &0.49     &3.20
&0.49 &0.23  
\enddata
\label{table3}
\end{deluxetable}

\end{document}